\documentclass[prd,onecolumn,superscriptaddress,nofootinbib,11pt]{revtex4-2}
\usepackage{graphicx}
\usepackage{epsfig}
\usepackage{bm}
\usepackage{amsfonts}
\usepackage{subfigure}
\usepackage{multirow}
\usepackage{float}
\usepackage{color}
\usepackage[usenames,dvipsnames]{xcolor}
\usepackage{ulem}
\usepackage{todonotes}
\usepackage{makecell}
\usepackage{enumitem}
\usepackage{amsmath}
\usepackage{amssymb}
\usepackage{caption}
\usepackage{verbatim}
\usepackage{array}
\usepackage[colorlinks=true,citecolor=Red,linkcolor=Green,urlcolor=Green]{hyperref}

\usepackage{amsmath,mathrsfs,amsbsy,color,graphicx,bm,amsthm,amsfonts}

\def\beq{\begin{equation}}
	\def\eeq{\end{equation}}
\def\ber{\begin{eqnarray}}
	\def\eer{\end{eqnarray}}
\def\benu{\begin{enumerate}}
	\def\eenu{\end{enumerate}}

\def\sq{\lower.25ex\hbox{\large$\Box$}}
\def \lleq {\lower0.9ex\hbox{ $\buildrel < \over \sim$} ~}
\def \ggeq {\lower0.9ex\hbox{ $\buildrel > \over \sim$} ~}

\usepackage{geometry}
\begin{document}

	\title{Dynamics of kinks in a traversable wormhole}
	\author{Tian-Chi Ma}\email{tianchima@buaa.edu.cn}
	\affiliation{Center for Gravitational Physics, Department of Space Science, Beihang University, Beijing 100191, China}
	\author{Hai-Qing Zhang}\email{hqzhang@buaa.edu.cn}
	\affiliation{Center for Gravitational Physics, Department of Space Science, Beihang University, Beijing 100191, China}
	\affiliation{Peng Huanwu Collaborative Center for Research and Education, Beihang University, Beijing 100191, China}

	\begin{abstract}
	We investigate the dynamics of spherically symmetric radial domain walls (or kinks) in a traversable Simpson-Visser wormhole. By solving the scalar field in a double-well potential, we find that the parameter $a$ has a strong impact on the kink dynamics: larger $a$ allows the kink to go through the throat back and forth, while smaller $a$ strongly confines the kink nearby the throat. In addition, each traverse of the kink through the throat is accompanied with the emission of scalar wave packets, resulting in a gradual decrease of the oscillation amplitude reminiscent of a damping oscillator. This oscillatory behavior between the two sides of the wormhole is in sharp contrast to its counterpart in compact objects, such as boson stars. Our findings uncover how wormhole geometry will influence the dynamics of topological defects and may provide new insights for distinguishing wormholes from ordinary compact objects.
		\end{abstract}
	
	\maketitle

\section{Introduction}
In field theories with spontaneous symmetry breaking, the causal disconnection between different regions of the system allows each region to select a different vacuum state. Topological defects then naturally emerge at the boundaries between these regions \cite{brandenberger1994topological,teo2017topological,kosterlitz2017nobel}. Depending on the class of symmetry being broken, different classes of defects, such as monopoles, cosmic strings, and domain walls may emerge \cite{pismen1999vortices,manton2004topological,bunkov2000topological}. Domain walls  (also commonly referred to as kinks) arise from the spontaneous breaking of $Z_2$ symmetry. They exhibit a step-like structure in which the scalar field undergoes a sharp transition from one vacuum to the other, leading to a localized concentration of energy density at the interface. Owing to their extended nature and relatively high tension, domain walls can induce significant gravitational effects and play an important role in early-universe cosmology, phase transitions, and nonperturbative field dynamics \cite{kibble1982phase, vilenkin1994cosmic, vachaspati2006kinks}. Domain walls are also often considered potential candidates for dark matter or other intriguing structures \cite{press1989dynamical,avelino2008dynamics,friedland2003domain}. Fundamental constants, such as particle masses and the fine-structure constant and etc., may vary upon crossing a domain wall \cite{kibble1976topology,vilenkin1982cosmic,vilenkin1985cosmic}. Since domain walls can alter fundamental physical constants, scientists have developed various experiments to detect them, including satellite synchronization \cite{derevianko2014hunting,roberts2017search,kalaydzhyan2017extracting,dai2022interaction}, electric dipole moment measurements \cite{stadnik2014searching}, magnetometry \cite{pustelny2013global,afach2021search}, acceleration induced by mass differences \cite{mcnally2020constraining}, and gravitational wave detectors \cite{hall2018laser,grote2019novel,jaeckel2021probing}. Therefore, studying domain walls or kinks in gravitational systems is a task of considerable importance.

The evolution of domain walls in flat spacetime has been extensively studied, including in both $2+1$ and $3+1$ dimensions. Reference \cite{kevrekidis2018planar} examines the dynamical evolution of both planar and radial kinks in Klein–Gordon models, with a particular focus on their existence and stability properties. The attractive interaction between a kink and an antikink has been analyzed in Ref. \cite{carretero2022kink}, where an effective model was constructed via the variational approximation, yielding results in agreement with those from the full dynamical model. In Ref. \cite{caputo2013radial}, the influence of radial perturbations on kink configurations is studied, and the possibility of extracting energy from the kink is explored. More recently, some studies have employed the gauge/gravity duality \cite{maldacena1999large,witten1998anti} to probe the statistical behavior of kinks in the strongly coupled regime of the boundary field theory \cite{li2023black,ma2025universal}. Additional related work can be found in references \cite{arodz1998expansion,dobrowolski2008construction,dobrowolski2009kink,gatlik2021modeling,gorria2004kink,shi2024topologically}.

In this context, an interesting question arises: compared with flat spacetime, what characteristic features does the kink exhibit in curved spacetime? Compact astrophysical objects, including black holes \cite{ruffini1971introducing,frolov2012black}, neutron stars \cite{lattimer2004physics,lattimer2014neutron}, boson stars \cite{jetzer1992boson,schunck2003general}, and wormholes \cite{hawking1988wormholes,morris1988wormholes,dai2020form}, serve as natural laboratories for studying the nonlinear behavior of fields under strong gravity. Compared with flat spacetime, the behavior of domain walls in curved geometry is significantly richer due to the background curvature, horizons, and the diverse structures of spacetime topology. In black hole backgrounds, scientists have conducted several studies on domain walls, including their stationary and transient behaviors. Studies have shown that black holes exhibit a repulsive effect on static thick domain walls \cite{moderski2003thick,morisawa2003thick,moderski2004reissner,moderski2006thick}. In addition, when a kink passes near a black hole, it displays behavior resembling ringdown modes \cite{ficek2018planar}. Recently, scientists have extended the study of kink dynamics to compact stars, such as neutron stars and boson stars. They found that compact stars can slow down the motion of kinks and may even produce a rebound-like effect on them \cite{caputo2024radial,ma2025radial}.

However, studies of kinks in wormhole backgrounds remain very limited, with current research mostly focusing on static configurations and perturbative analyses \cite{sushkov2002wormholes,bizon2021sine,diaz2023kinks,waterhouse2019phi,goulart2017phantom}. A wormhole refers to a tunnel-like spacetime structure that links two distant or otherwise disconnected regions. The earliest wormhole solution, known as the Einstein–Rosen bridge, was proposed in \cite{einstein1935particle}. Since this type of wormhole is non-traversable, it was long considered merely a mathematical construction rather than a physically realizable entity. Many years later, Ellis obtained a new wormhole solution, and Morris and Thorne subsequently proved that the Ellis wormhole is traversable, such that matter or light signals can safely pass through the wormhole without encountering horizons or singularities \cite{ellis1973ether,morris1988wormholes2}. The existence of traversable wormholes requires matter that violates classical energy conditions, known as ``exotic matter", whose negative energy density can support the wormhole throat against collapse \cite{chianese2017characterising,di2017spin}. For further studies on wormholes, see Refs. \cite{battista2024generalized,dai2018new,de2021reconstructing,de2021testing,de2020general,dai2019observing,bambi2021astrophysical,de2025can,de2021epicyclic,de2023static}.

Moreover, in traversable wormholes, the two asymptotically flat regions may be interpreted as separate universes. In such setups, a kink located in one universe could, in principle, travel through the wormhole throat and enter the other universe.
The arrival of such a high-tension defect could substantially alter the local gravitational environment, disrupt astrophysical systems, or leave observable imprints in cosmology. Therefore, studying the evolution of kinks in wormhole spacetimes has significant theoretical and even practical importance. To our knowledge, the present work constitutes the first detailed study of the spherically symmetric radial kink dynamics in the background of a traversable wormhole.

In this paper, we investigate the time evolution of radial domain walls in the background of the Simpson-Visser two-way traversable wormholes. In Sec. \ref{sect2}, we introduce the Simpson-Visser spacetime and the spherically symmetric scalar-field model with a double-well potential. In Sec. \ref{sect3}, we solve the scalar field equation to investigate how the width of the wormhole throat affects the dynamical evolution of the kink, as well as the mechanism by which the kink transfers energy to the background. In addition, by studying the amplitude of the kink's oscillation and the energy of the kink, we find that the oscillation of kink is similar to a damping oscillator. Finally, we draw conclusions and discussions in Sec. \ref{sect4}. Throughout this work, we adopt geometric units $G=c=1$ and the metric signature $(-,+,+,+)$.

\section{Theoretical framework}\label{sect2}

\subsection{Simpson-Visser spacetime and wormhole configurations}
The Simpson-Visser spacetime is described by the line element
presented in \cite{simpson2019black}, 
\begin{align}\label{eq:WHmetric}
	ds^{2} & =-A\left(r\right)dt^{2}+\frac{1}{A\left(r\right)}dr^{2}+\left(r^{2}+a^{2}\right)\left(d\theta^{2}+\sin^{2}\theta d\varphi^{2}\right), \\
	A\left(r\right) & =1-\frac{2M}{\sqrt{r^{2}+a^{2}}},
\end{align}
where $M$ is the ADM mass while $a$ is the characteristic parameter of this spacetime. The coordinate range of this spacetime is
\begin{equation}
    t\in(-\infty,+\infty);\qquad
	r\in(-\infty,+\infty);\qquad 
	\theta\in [0,\pi];\qquad
	\phi\in[0,2\pi).
\end{equation}

By changing the parameter $a$, this spacetime can exhibit four different shapes: (1) The ordinary Schwarzschild spacetime for $a=0$; (2) A regular black hole geometry with a one-way {\it spacelike} throat for $0<a<2M$; (3) A one-way wormhole geometry with an extremal {\it null} throat for $a=2M$; (4) A canonical traversable wormhole geometry with a two-way {\it timelike} throat for $a>2M$. In this work, we will focus on the last scenario by studying the time evolution of radial kinks in the two-way traversable wormholes. Therefore, in the following we will restrict our analysis to the case of $a>2M$.

To better illustrate the geometry and properties of the two-way traversable wormhole, we take advantage of the embedding diagrams. Specifically, we fix the time $t=\text{const.}$ and $\theta=\pi/2$, and then embed a two-dimensional spatial hypersurface of the wormhole spacetime into a three-dimensional Euclidean space. This allows the intrinsic properties of the wormhole spacetime to be intuitively displayed.  
By introducing the cylindrical coordinates $(\rho,\varphi,z)$, the metric on this two-dimensional hypersurface becomes
\begin{align}
	 ds^2 = \frac{1}{A\left(r\right)}d r^2 + (r^2+a^2)d\varphi^2 
	= d \rho^2 + dz^2 + \rho^2 d \varphi^2   \,.
\end{align}
in which $\rho$ and $z$ are
\begin{equation} \label{formula_embedding}
	\rho(r)= \sqrt{r^2+a^2} ,\;\;\;\;\;\;\;\;\;\;   z(r) = \int_0^r  \sqrt{\frac{1}{A(r)}  -   \left( \frac{d \rho}{d r} \right)^2    }     d r \;.
\end{equation}

\begin{figure}[h]
	\centering
	\includegraphics[trim=3.3cm 9cm 2cm 9.5cm, clip=true, scale=0.49]{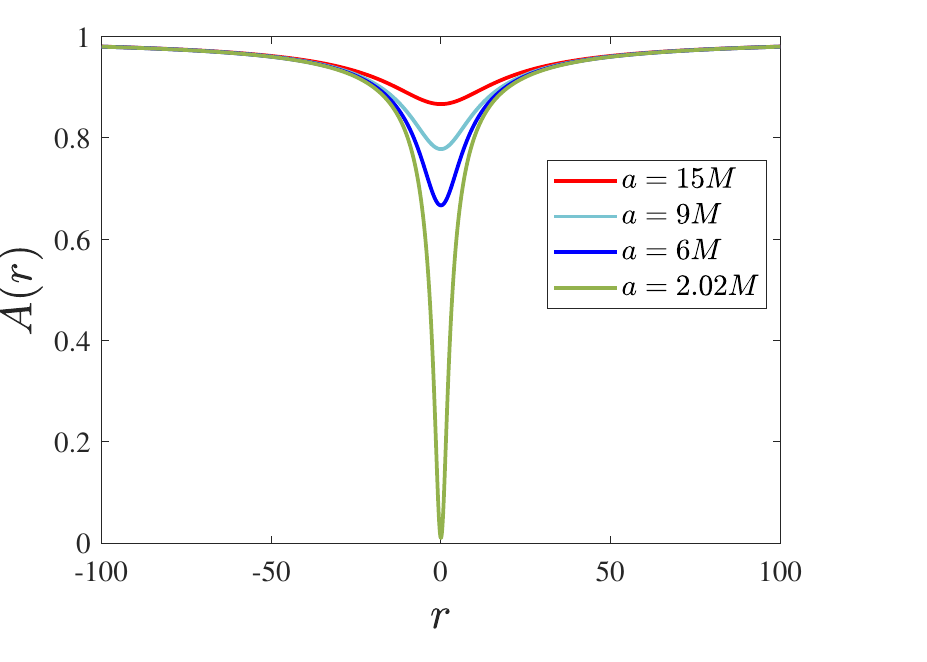}
	\put(-225,147){(a)}~
	\includegraphics[trim=3.3cm 9cm 2cm 9.5cm, clip=true, scale=0.49]{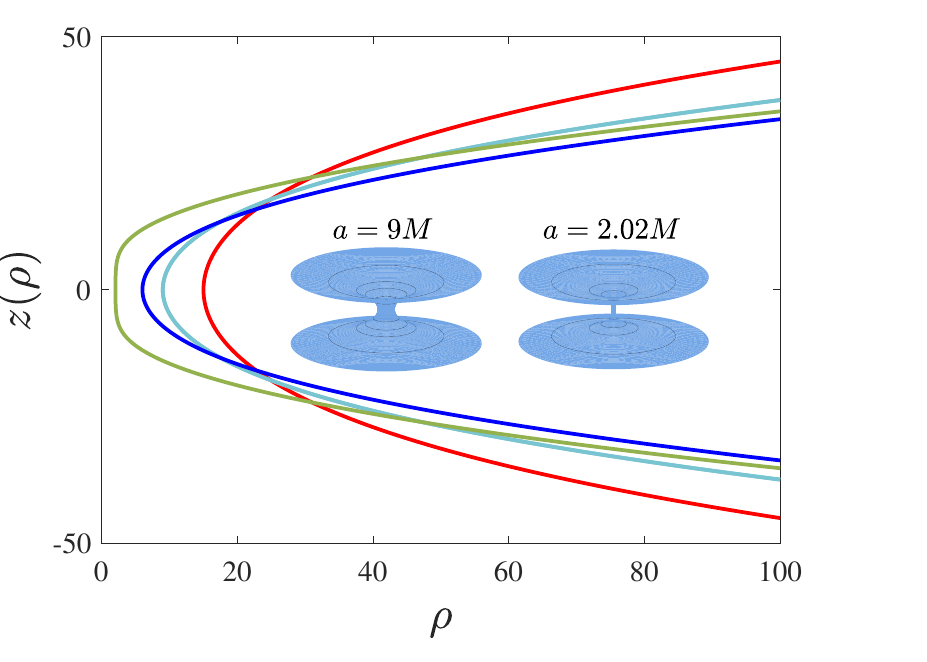}
	\put(-225,147){(b)}~
	\caption{The metric function $A(r)$ (in panel (a)) and the embedding function $z(\rho)$ (in panel (b)) with four different values of $a$, i.e., $a=15M$, $a=9M$, $a=6M$ and $a=2.02M$. The inset plot in panel (b) shows the three-dimensional configurations of the embedding function $z(\rho)$. It can be seen that as $a$ is smaller the wormhole throat is narrower. 
	}\label{fig:ArAndzr}
\end{figure}

In panels (a) and (b) of Fig.\ref{fig:ArAndzr}, we present a set of metric functions $A(r)$ and the embedding diagrams $z(\rho)$ with different values of $a$. It is observed from Fig.\ref{fig:ArAndzr}(a) that for larger values $a$, the metric function $A(r)$ exhibits a shallower valley, and correspondingly, in Fig.\ref{fig:ArAndzr}(b) the embedding diagram has a wider throat. See for instance the red curves with $a=15M$. On the contrary, for smaller values of $a$, the metric function $A(r)$ has a deeper valley, corresponding to a narrower throat in Fig.\ref{fig:ArAndzr}(b). Please refer to the green curves with $a=2.02M$. The inset plot in Fig.\ref{fig:ArAndzr}(b) shows the three-dimensional configurations of the embedding function $z(\rho)$ with $a=9M$ and $a=2.02M$. It can be intuitively seen that the embedding diagram of the wormhole is divided into two symmetric halves along $z=0$ plane, resembling two funnels that are joined at their narrowest ends. The upper and lower halves correspond to $r>0$ and $r<0$, respectively. The narrowest part is referred to as the throat of the wormhole corresponding to $z=0$.

\subsection{Model of radial kinks}
In order to study the dynamics of kinks in the background of the two-way traversable wormholes, we consider a real scalar field $\phi$ with the action
\begin{equation}\label{S}
	S = \int d^4 x \sqrt{-g} \left[ \frac{1}{2} \, g_{\mu \nu}
	{\nabla}^{\mu} \phi {\nabla}^{\nu} \phi + V(\phi) \right] ,
\end{equation}
where we will adopt the metric $g_{\mu\nu}$ as in Eq. \eqref{eq:WHmetric}, and the potential $V(\phi)$ as a double-well potential
\ber V(\phi) = \frac{1}{4} (\phi^2 - 1)^2, \eer 
which has minimums at $\phi=\pm 1$. Therefore, after breaking the $Z_2$ symmetry in the action \eqref{S}, the scalar field $\phi$ will settle down to $\phi=+1$ or $\phi=-1$, which will in turn form the solutions of kinks \cite{kibble1976topology,kibble1980some,zurek1985cosmological,ma2024asymmetric}. The equation of motion for the scalar field $\phi$ reads
\begin{equation}\label{eq:phimove}
	-\frac{1}{\sqrt{-g}} \,\partial_{\mu} \left( \sqrt{-g} \,g^{\mu \nu}
	\partial_{\nu} \phi \right) + \frac{\delta V(\phi)}{\delta \phi} = 0 .
\end{equation}
 
In this work, we are interested in the time evolution of the spherically symmetric kinks, therefore, in the background of spherical geometry the Eq. \eqref{eq:phimove} can be re-expressed as 
\begin{align}\label{eq:phimove2}
	&\partial_t^2 \phi - \frac{1}{r^2+a^2} \,  A(r) \, \partial_r \left( (r^2+a^2) A(r) \, \partial_r \phi \right)  
	- \frac{1}{(r^2+a^2) \sin \theta} \,  A(r) \, \partial_{\theta} \Big( \sin \theta \, \partial_{\theta} \phi \Big)\nonumber \\	&- \frac{1}{(r^2+a^2) \sin^2 \theta} \,  A(r) \, {\partial^2_{\varphi}  \phi } 
	+  A(r) \phi^3 -  A(r) \phi = 0 .
\end{align}
Due to the spherical symmetry of the system, the polar and azimuthal angular derivative terms in the equation do not play a role. Therefore, we can simplify the configurations of the kink to only depend on time $t$ and the radial direction $r$. And the final equation simplifies to
\begin{align}\label{eq:phimove3}
	&\partial_t^2 \phi - \frac{1}{r^2+a^2} \,  A(r) \, \partial_r \left( (r^2+a^2) A(r) \, \partial_r \phi \right)   
	+  A(r) \phi^3 -  A(r) \phi = 0 .
\end{align}
Hence we dub the kink solutions in this case as ``{\it radial kinks}".

\section{Results: dynamics of radial kinks in the Simpson-Visser wormholes}\label{sect3}
\subsection{Initial configurations and numerical schemes}
In this section we will study the evolution of kinks in the background of the Simpson-Visser wormholes. The dynamics of kinks are governed by the Eq. \eqref{eq:phimove3}. The initial configuration of the kink can be given by \cite{Weinberg:1992hc},
\begin{equation}
	\label{eq:phiini1}
	\phi(0,r)= \tanh \left[ \frac{r-r_k(0)}{\sqrt{2 (1-v^2)}} \right] ,
\end{equation}
where $r_k(0)$ is the initial position of the kink at $t=0$ while $v$ represents the initial velocity of the kink. Throughout the paper, we always set $v=0$. This means that within the framework of the field model described by Eq. \eqref{eq:phimove3}, we assume initial conditions of the kink with a given radius and zero initial velocity. The position of the kink $r_k$ is defined at the transition that $\phi$ changes its sign, i.e., $\phi(t,r_k)=0$. 

In numerics, we set the radial direction $r \in [-100,100]$ and we discretize it into $5000$ grid points. At the boundary of $r=\pm 100$, we impose the Dirichlet boundary condition of the scalar field $\phi(t,\pm 100)=\pm 1$. This requirement is consistent with the initial condition. Besides, we employ the fourth-order Runge-Kutta method in the time direction with a time step of $\Delta t=0.02$. In the radial direction $r$, we used the sixth-order difference method with $\Delta r=0.04$. To check the numerical stability, we also tested other time and spatial step sizes, such as $\Delta t=0.04$, $\Delta r=0.08$ or $\Delta t=0.01$, $\Delta r=0.02$, and obtained consistent results.

\subsection{Traverse of the radial kinks through the wormhole}

\begin{figure}[htbp]
	\centering
	\includegraphics[trim=3.3cm 8.3cm 2cm 9.5cm, clip=true, scale=0.49]{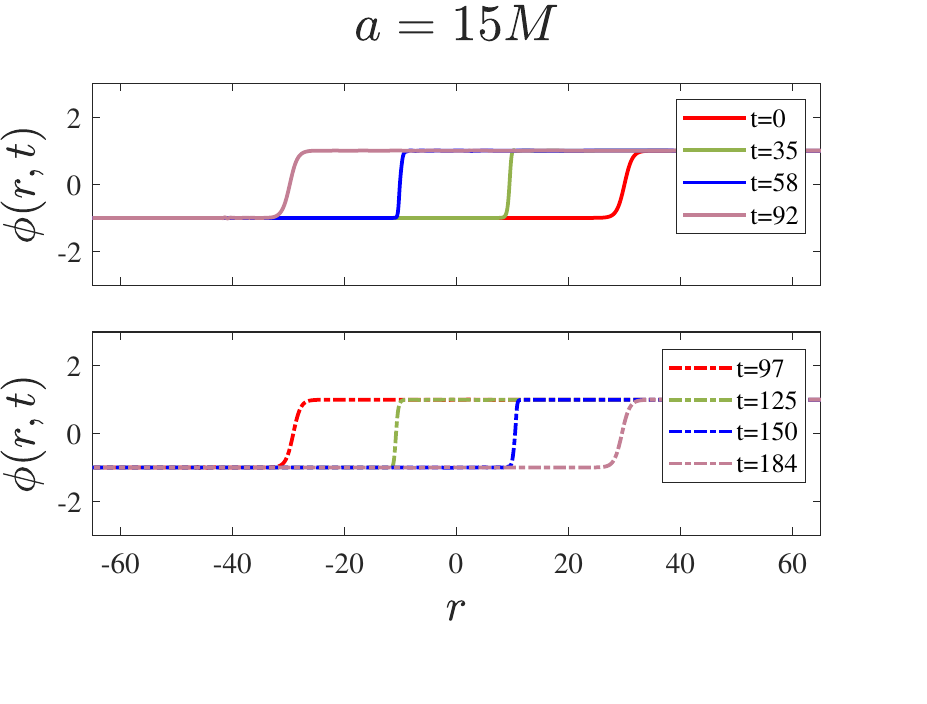}
	\put(-225,147){(a)}~
	\includegraphics[trim=3.3cm 8.3cm 2cm 9.5cm, clip=true, scale=0.49]{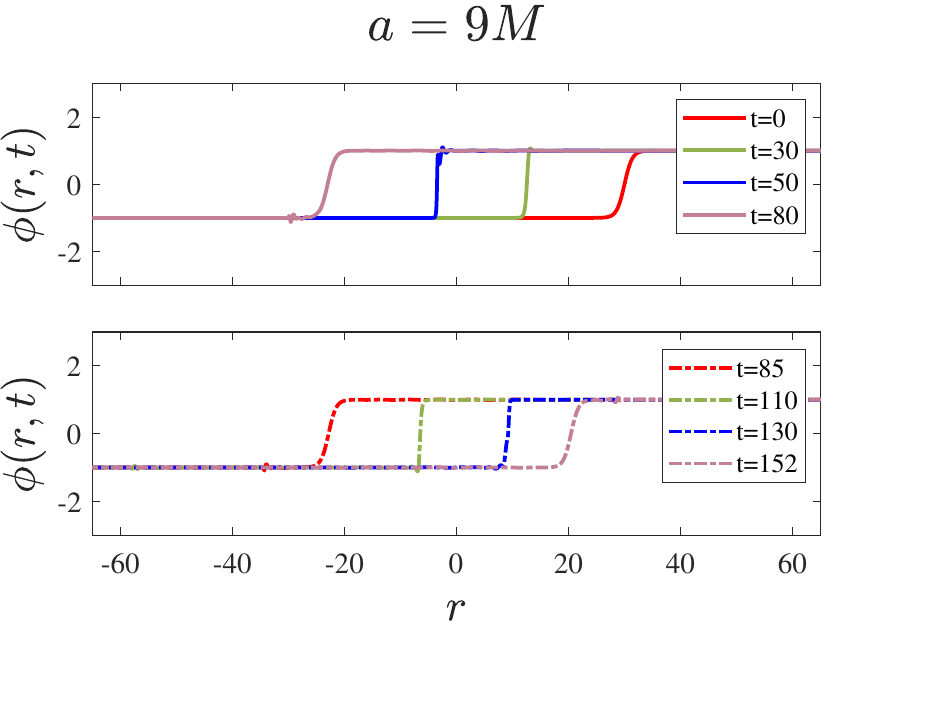}
	\put(-225,147){(b)}~\\
	\includegraphics[trim=3.3cm 9.5cm 2cm 9.5cm, clip=true, scale=0.49]{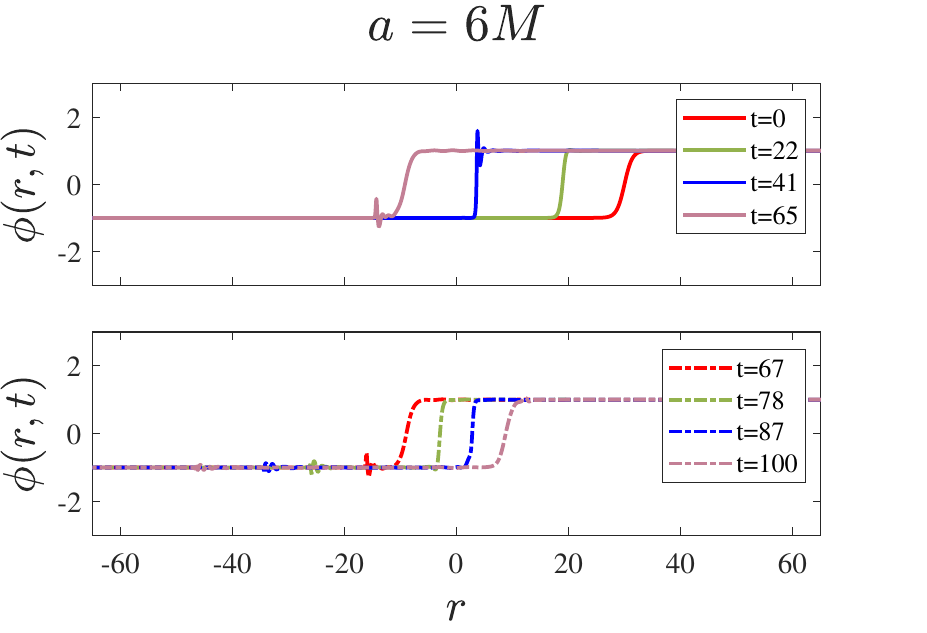}
	\put(-225,130){(c)}~
	\includegraphics[trim=3.3cm 9.5cm 2cm 9.5cm, clip=true, scale=0.49]{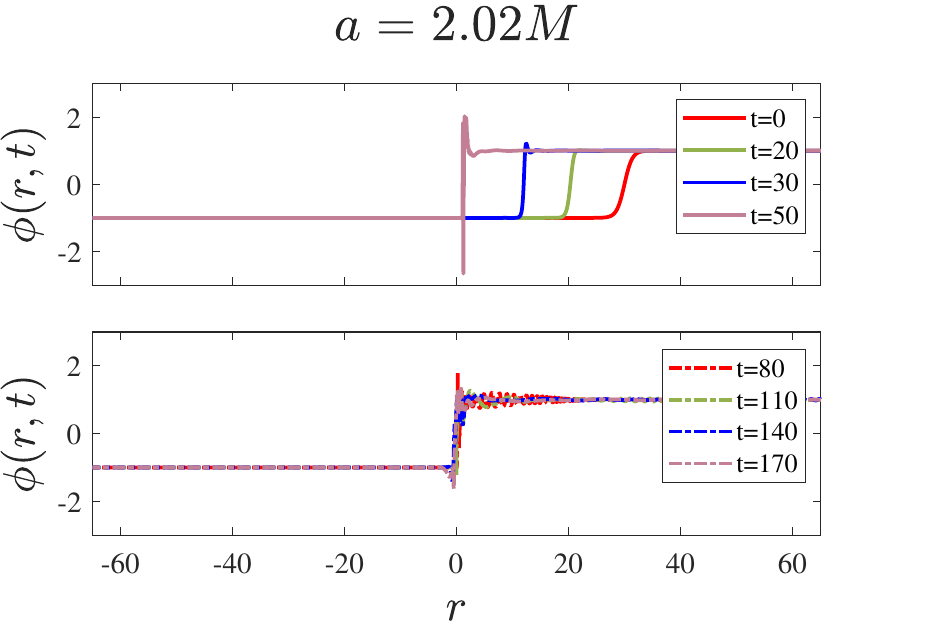}
	\put(-225,130){(d)}~
	\caption{Time evolution of the kinks in the wormhole background with four different values of $a$. In each panel, the scalar field shares the same initial configuration with the kink position at $r_k(0)=30$. Each panel is divided into two parts, in which the solid lines (in the upper part) represent the kinks moving leftwards, while the dashed lines (in the lower part) represent the kinks moving rightwards. However, in the panel (d) we can see that the dashed lines in the lower part are confined nearby the throat of the wormhole. 
	}\label{fig:KinkfourA}
\end{figure}

Fig.\ref{fig:KinkfourA} shows the time evolution of the radial kink with four different values of the parameter $a$ such that $a=15M$, $a=9M$, $a=6M$ and $a=2.02M$, respectively. (In numerics we have set $M=1$.) In each panel, we use the same initial scalar field configuration, with the kink initially located at $r=30$. From the figure, it is clear that for larger values of $a$ (e.g., $a=15M$), the kink starts moving leftward from $r=30$. Later it traverses through the throat of the wormhole, reaches a maximum displacement at $r\approx -29.83$ ($t=92$) on the left side of the wormhole, and then bounces back to move rightward reaching up to $r\approx 29.47$ ($t=184$). It should be noted that, after the time $t=184$ the kink bounces back again to move leftward and repeats its evolution back and forth. The back-and-forth motion of the kink across the wormhole's two horizons makes it an excellent candidate as a mechanism for retrieving information from the opposite side of the wormhole.

However, for smaller values of $a$ (i.e., narrower throats), the kinks travel a shorter distance when they propagate across the throat although they have the same initial conditions as $a=15M$. Please refer to the panels (b), (c) and (d) in Fig.\ref{fig:KinkfourA}. In particular, as $a$ is very close to the critical value $a=2M$ shown in the panel (d) of Fig.\ref{fig:KinkfourA} with $a=2.02M$, the kink's motion will be confined nearby the throat of the wormhole, rather than going through the wormhole. This is consistent with the fact that as $a\leq2M$, the spacetime has a null or one-way spacelike throat. It is clear from Fig.\ref{fig:KinkfourA} that as $a$ is smaller, the narrower throat will hinder the traverse of the kinks. This can be seen from the more apparent ripples of wave packets arising from panel (b) to panel(d), which takes more energy away from the kink and then results in the decreasing amplitude of the bounce.

\subsection{Wave packets shedding and energy transfer}\label{shed}

In the previous subsection we have observed the ripples appeared in the scalar fields as shown in the panels (b)-(d) in Fig.\ref{fig:KinkfourA}. These ripples usually turn out as they are close to the throat of the wormhole. Besides, our previous results also show that the amplitude of the kink's back-and-forth bounce decreases over time, indicating that the energy of the kink reduces over time. In order to investigate this phenomenon, we will perform a further study in the following by the case of $a=9M$ as an example.

\begin{figure}[htbp]
	\centering
	\includegraphics[trim=3.2cm 8.3cm 3cm 9.5cm, clip=true, scale=0.39]{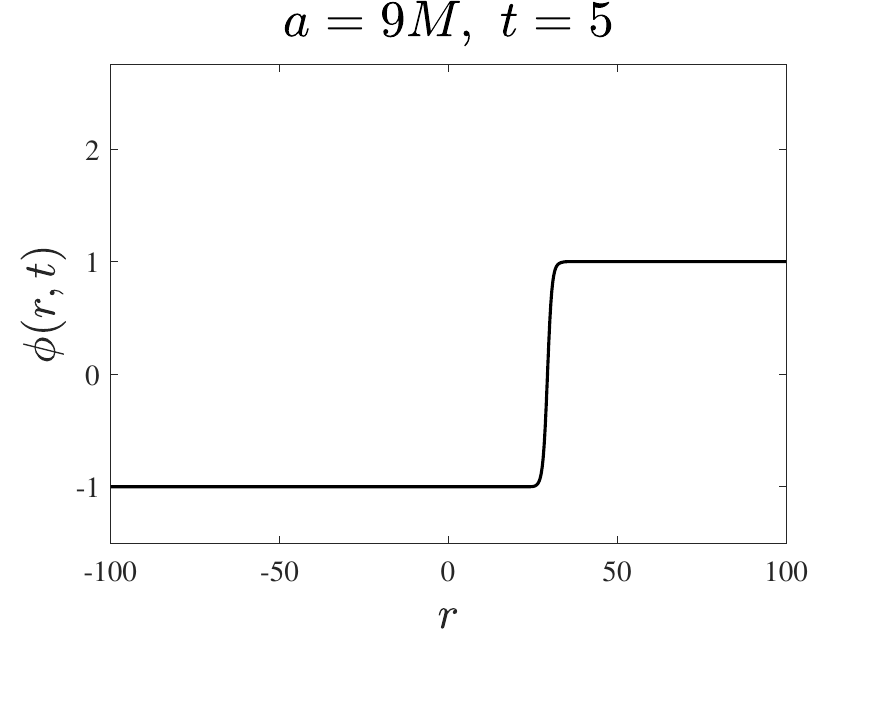}
	\put(-167,122){(a)}
	\includegraphics[trim=3.2cm 8.3cm 3cm 9.5cm, clip=true, scale=0.39]{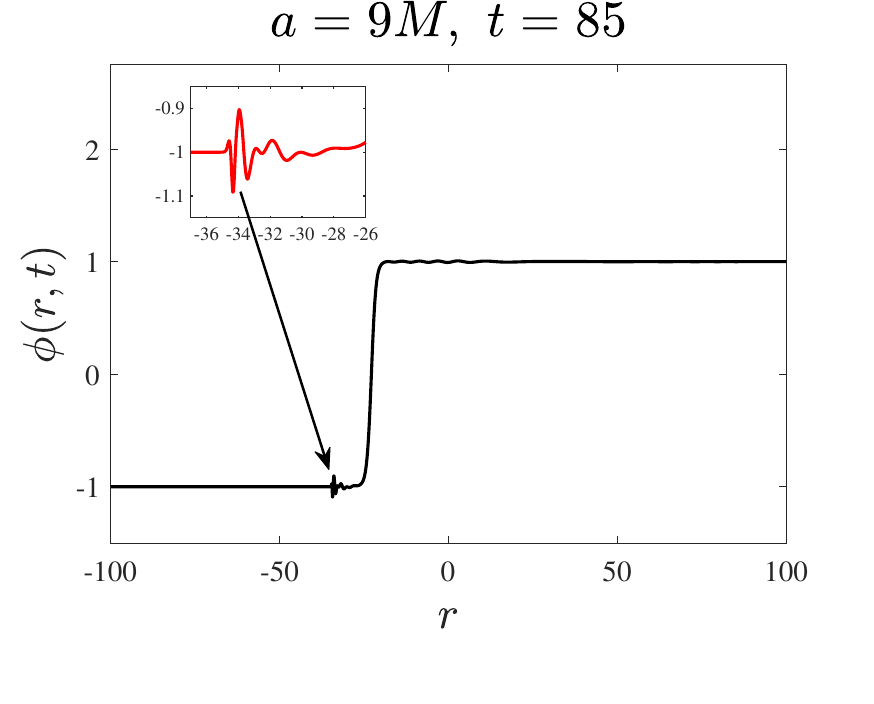}
	\put(-167,122){(b)}
	\includegraphics[trim=3.2cm 8.3cm 3cm 9.5cm, clip=true, scale=0.39]{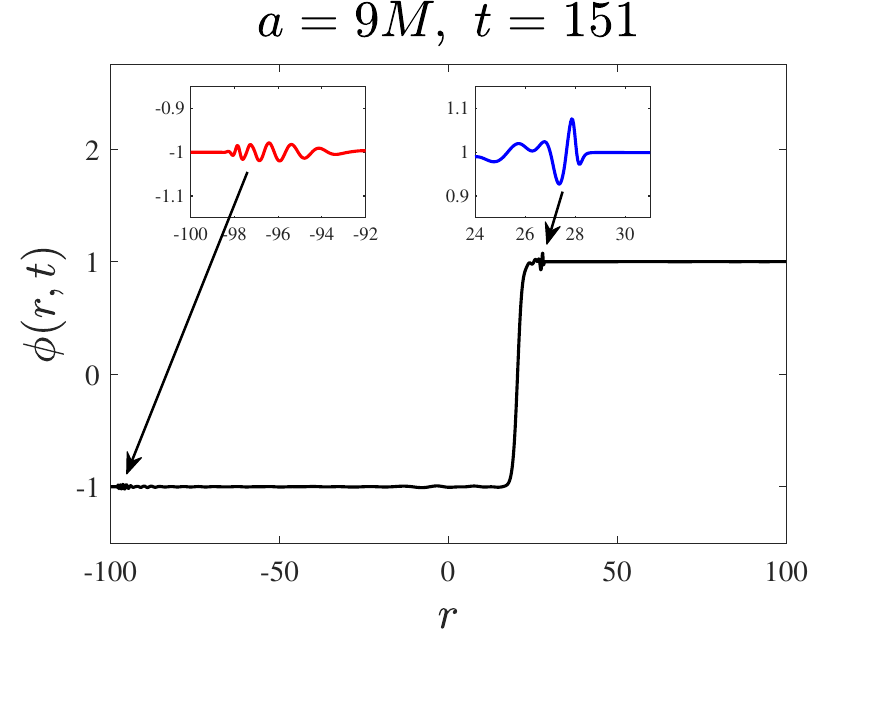}
	\put(-167,122){(c)}~\\
	\includegraphics[trim=3.2cm 9.5cm 3cm 9.5cm, clip=true, scale=0.39]{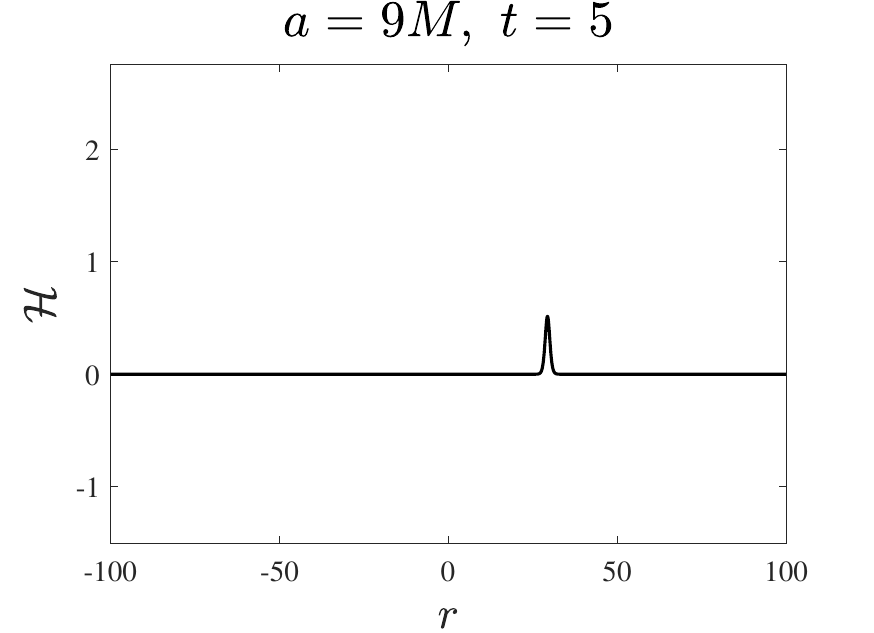}
	\put(-167,108){(d)}
	\includegraphics[trim=3.2cm 9.5cm 3cm 9.5cm, clip=true, scale=0.39]{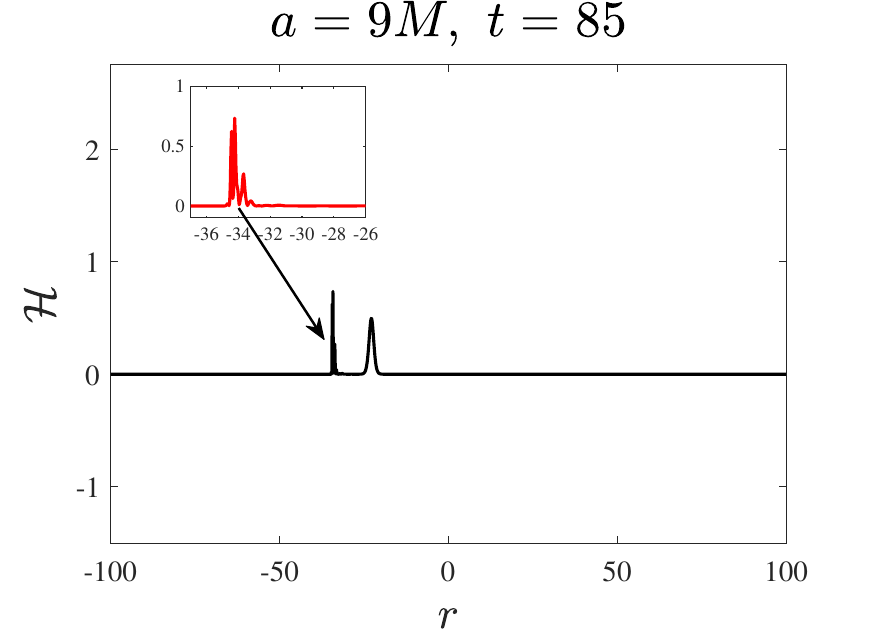}
	\put(-167,108){(e)}
	\includegraphics[trim=3.2cm 9.5cm 3cm 9.5cm, clip=true, scale=0.39]{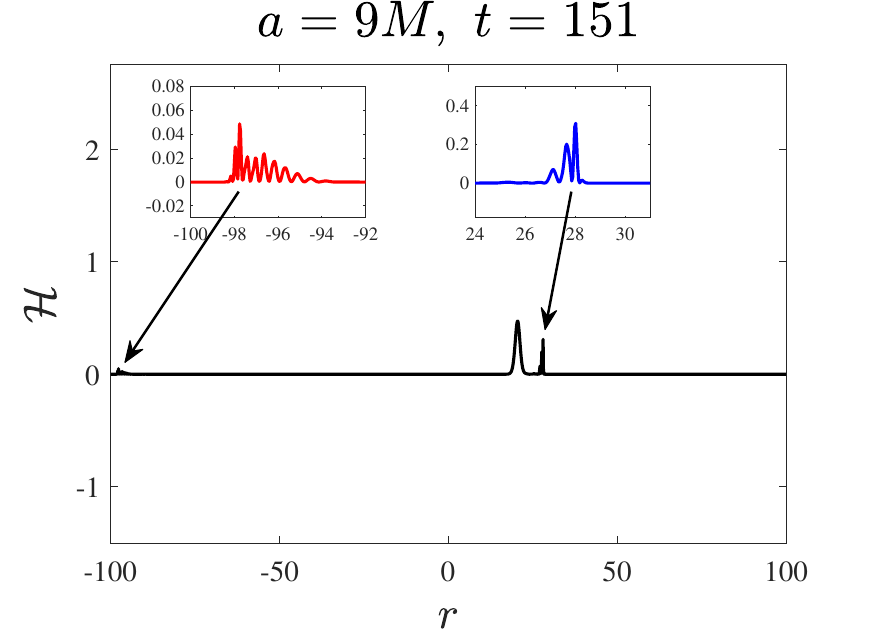}
	\put(-167,108){(f)}
	\caption{Configurations of the scalar field $\phi$ and its corresponding energy density $\mathcal{H}$ at different times, i.e., $t=5, 85$, and $151$. The initial position of the kink is $r_k(0)=30$ with the parameter $a=9M$. The inset plots are the enlarged version of the wave packets or the energy densities as the arrows indicate.  
	}\label{fig:picKinkDensity}
\end{figure}

The energy density of the system $\mathcal{H}$ on a spacelike hypersurface can be obtained as 
\begin{align}\label{Henergy}
\mathcal{H}=n^{\mu} \xi^{\nu} T_{\mu \nu}.
\end{align}
Here, we assume the spacelike hypersurface is a hypersurface $\Sigma_t$ with constant time. Therefore, $n^{\mu}=(1/\sqrt{A(r)}, 0, 0, 0)$ is the unit normal vector to the hypersurface, $\xi^{\mu}=(1, 0, 0, 0)$ is the timelike Killing vector and $T_{\mu \nu}$ is the energy-momentum tensor given by
\begin{equation}
	\label{T}
	T_{\mu \nu} = \nabla_{\mu} \phi \, \nabla_{\nu} \phi - g_{\mu \nu} \, {\cal L} ,
\end{equation}
Therefore, from the action of scalar field \eqref{S} the energy density finally becomes
\begin{align}
	\label{eq:HDen}
\mathcal{H}=\frac{1}{\sqrt{A}}\left[ \frac{1}{2} \, (\partial_t \phi)^2 + \frac{1}{2} A^2\, (\partial_r \phi)^2 + A V(\phi)\right].
\end{align}

In the panels (a)-(c) of Fig.\ref{fig:picKinkDensity}, we show the configurations of the kinks at three different times with $a=9M$. Meanwhile, in the panels (d)-(f) of Fig.\ref{fig:picKinkDensity}, we exhibit their corresponding energy densities $\mathcal{H}$, respectively. At $t=5$ (panels (a) and (d)), the kink starts from its initial location and moves towards the wormhole throat, but has not yet arrived at it. It is observed that at this time there are no ripples in the scalar fields. Besides, the energy density peaks at the kink location, while it is vanishing at other positions. This is because the scalar field $\phi$ exhibits a step-like profile at the kink position, which contributes the most excited energy to the kink. In contrast, at other positions the scalar field is flat, which stays in its lowest energy state \cite{allen1979microscopic,Shi:2023qnb}. 

At $t=85$ (panels (b) and (e)), the kink already passed through the throat once. It is observed that the field configuration and energy density at $t=85$ differ significantly from those at $t=5$. In panel (b), aside from the kink, a wave packet is present at around $r\approx -34$ (see the inset plot for the enlarged configurations).  Therefore, the energy of the kink will decrease after it passes through the throat.  Correspondingly, in panel (e) of Fig.\ref{fig:picKinkDensity}, except for a peak of the energy density at the location of the kink, there also exists an excitation of energy density at the positions of the excited wave packet. 

Then, the kink will bounce off at around $r\approx -23$ and move rightwards to the throat again. However, the shedding wave packet in panel (b) will continue moving leftwards until it hits the left boundary $r=-100$. This phenomenon is captured in the panels (c) and (f) in Fig.\ref{fig:picKinkDensity} at time $t=151$. In this case, the kink already passed through the throat twice. There appears another new generated wave packet at around $r\approx 28$, which is represented by the blue inset plot in panel (c). The wave packet in red inset plot in panel (c) is from the wave packet in panel (b), which has propagated leftwards to $r\approx -97$ at time $t=151$. The corresponding energy densities of the kinks and the two wave packets are shown in panel (f). 
Later, similar behaviors will go on in the subsequent time evolution. That is, each time the kink passes through the wormhole throat, there will emit a wave packet, transferring the energy to the background. Consequently, the amplitude of the bouncing will gradually decrease, as shown in the Fig.\ref{fig:picEkR2}(a) in next section.

The importance of the back-and-forth bouncing of the kinks between the wormhole's two sides is that we may infer the geometry of the wormhole by observing the damping oscillations or the wave packet shedding of the kink in future. Further, the bouncing may bring back the information of the other side of the wormhole to the observers who only live in this side of the wormhole.

\subsection{Damping oscillations and energy loss}
\label{damp}

\begin{figure}[htbp]
	\centering 
	\includegraphics[trim=3.1cm 9.5cm 2.5cm 9.5cm, clip=true,width=.5\textwidth]{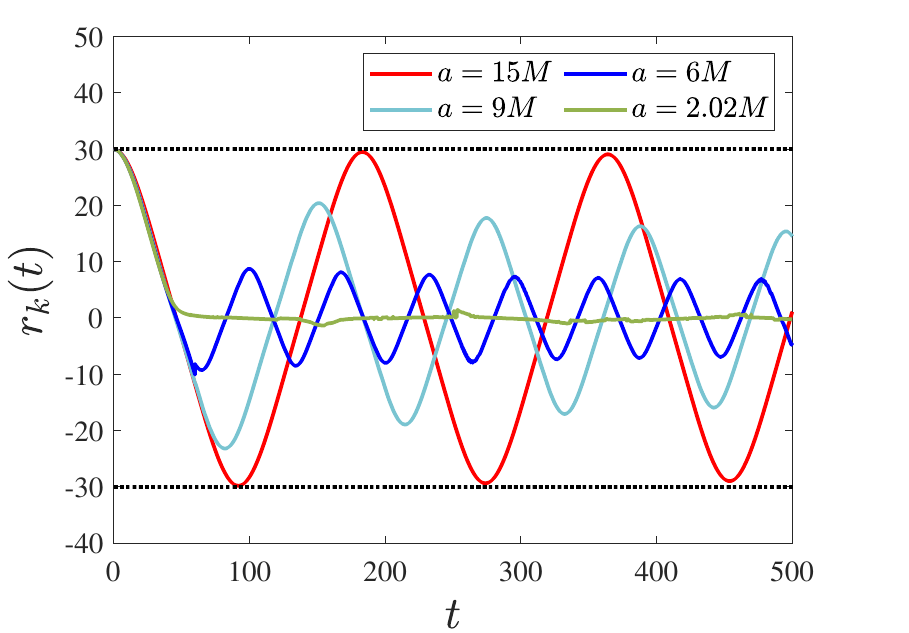}
	\put(-240,160){(a)}
	\includegraphics[trim=3.1cm 9.5cm 2.5cm 9.5cm, clip=true,width=.5\textwidth]{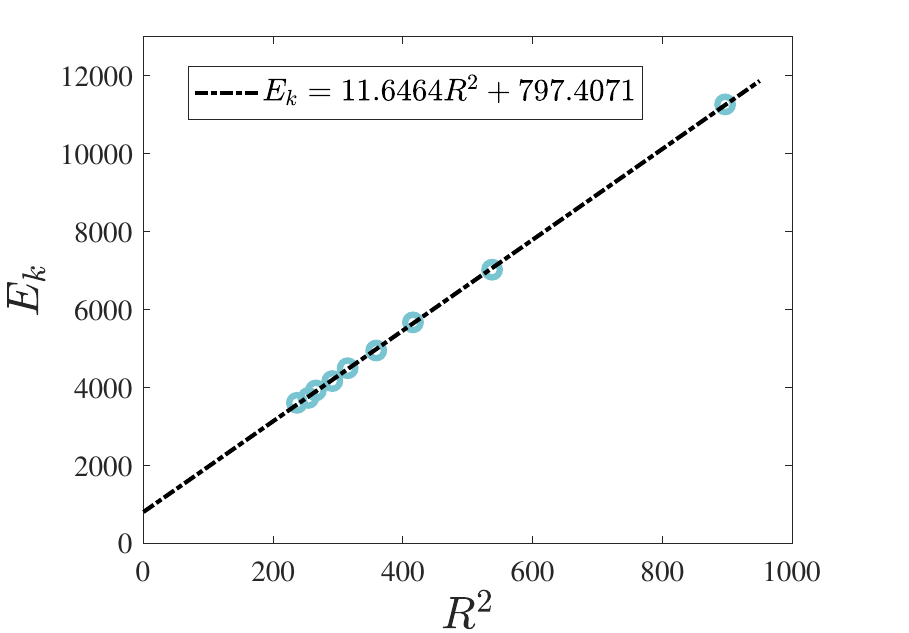}
	\put(-240,160){(b)}
	\caption{(a) Time evolution of the kink's position $r_k(t)$ with four different values of the parameter $a$. In order to better illustrate the evolutions of the kink position, two dotted lines at $r_k=\pm30$ have been added as a reference; (b) Linear relation between the kink's energy and the amplitude's square. The fitting line (dot-dashed) matches the numerical data very well.}
	\label{fig:picEkR2}
\end{figure}

In order to gain a full understanding of the oscillations of the kinks traversing between the two sides of the wormhole, we exhibit the time evolutions of the kink positions $r_k(t)$ in Fig. \ref{fig:picEkR2}(a). It is seen that the amplitudes of the oscillations decrease over time. In particular, for smaller $a$ the reduction in oscillation amplitude becomes more pronounced. This observation is consistent with the embedding diagrams shown in Fig.\ref{fig:ArAndzr}, where smaller values of $a$ correspond to narrower wormhole throats, indicating that narrower throat will prevent the kinks to go through.  

It should be noted that as $a$ is close to the critical value such as $a=2.02M$, kinks will eventually confined nearby the throat as shown in Fig.\ref{fig:KinkfourA}(d). Therefore, it is hard to track the position of the kink for the case of $a=2.02M$, since many complex oscillatory modes will superimpose on the kink. Hence, in Fig.\ref{fig:picEkR2}(a), the curve corresponding to $a=2.02M$ takes the averagely approximate range of the kink’s location. \footnote{The full time animations of the kink evolution can be found in this \href{https://bhpan.buaa.edu.cn/link/AA26E0EBA8C19546C699D99CB8B155200E}{link}.}
The reduced amplitudes in Fig.\ref{fig:picEkR2}(a) reminds us that the oscillations of the kinks resemble damping oscillators, therefore, it is indispensable to investigate the relations between the amplitude of the oscillations and the energy of the kinks in order to see whether it is consistent with those of damping oscillators.

The energy of the kink can be obtained by integrating the scalar-field energy density $\mathcal{H}$ in Eq.\eqref{eq:HDen} over the region occupied by the kink,
\begin{equation}
	E_k = \int_{V_D} \sqrt{h} \,\mathcal{H} \,\, d^3 x ,
\end{equation}
where $V_D$ denotes the spatial volume containing the kink only, excluding the detached wave packet, and $h$ is the determinant of the induced metric on the spacelike hypersurface. Substituting the metric functions and making use of Eq. \eqref{eq:HDen}, we obtain
\begin{equation}
	E_k = \int_{r_D} \int_0^{\pi} \int_0^{2\pi} \frac{1}{A} \,\left[ \frac{1}{2} \, (\partial_t \phi)^2 + \frac{1}{2} A^2\, (\partial_r \phi)^2 + A V(\phi)\right] \,\, (r^2+a^2) \sin \theta d  r d \theta d \varphi  \, ,
\end{equation}
where $r_D$ denotes the radial interval occupied by the kink. Taking into account of the spherical symmetry of the system, we integrate out the angular variables and obtain
\begin{equation}
	\label{E2-r3}
	E_k = 4 \pi \int_{r_D} \frac{1}{A} \left[ \frac{1}{2} \, (\partial_t \phi)^2 + \frac{1}{2} A^2\, (\partial_r \phi)^2 + A V(\phi)\right](r^2+a^2) d  r  .
\end{equation}

Without loss of generality, we take $a=9M$ as an example in Fig.\ref{fig:picEkR2}(b) and plot the kink energy $E_k$ against the square of its oscillation amplitude $R^2$ during the time interval $t\in[0,500]$.\footnote{In damping oscillators, the relation between the time-dependent energy and the amplitude is $E(t)=\frac{1}{2}k R(t)^2$ where $k$ denotes the spring constant and $R(t)$ is the time-dependent amplitude.}  We find that the kink energy exhibits a linear dependence on the square of the amplitude,
\begin{equation}
	\label{eq:Enfit}
	E_k(t)=c_1 R(t)^2+c_2  ,
\end{equation}
where $c_1\approx 11.6464$ and $c_2\approx 797.4071$ are fitting coefficients. Please note that the fitting formula Eq.\eqref{eq:Enfit} has an extra constant term compared to that in damping oscillators. This is because the constant term $c_2$ is the residual energy of the kink when it does not oscillate anymore and stay rest at the wormhole throat $r=0$. Consequently, one can obtain an intuitive understanding of the dynamics of kinks. Each time the kink traverses the wormhole throat, part of its energy is radiated away in the form of a wave packet. The resulting energy loss leads to a progressive decrease in the oscillation amplitude of the kink. In this sense, the kink behaves analogously to a damping oscillator, with the emitted wave packets playing the role of an effective dissipation mechanism.

\section{Conclusions and Discussions}\label{sect4}
In this work, we studied the evolutions of the radial kinks in the background of a traversable Simpson-Visser Wormhole. We have uncovered a long-lived oscillatory behavior of radial kinks which can traverse back-and-forth through the wormholes. Our results show that the parameter $a$ has a decisive influence on the motion of the radial kink. That is, larger values of $a$ lead to a wider throat, allowing the kink to oscillate between the two asymptotically flat regions with a larger amplitude, whereas smaller values of $a$ confine the kink to oscillate with a smaller amplitude. In particular, as $a$ approaches the critical value $a=2M$, the throat becomes an extremal null throat, which confines the kink inside the throat. It is worth pointing out that the oscillating behavior of kinks between the two sides of the wormhole may help people to extract the information from the other side of the wormhole.

Due to the fact that the topology of the wormhole differs from that of ordinary compact stars, a single radial kink is protected by the wormhole topology and does not disappear at late times. Each time the radial kink passes through the wormhole throat, it transfers energy to the background in the form of emitted wave packets. Therefore, it reduces its own energy and causes its oscillation amplitude to gradually decrease, which resembles a damping oscillator. This only drives the kink to oscillate within a progressively smaller region near the throat, while the kink itself persists throughout the evolution. In contrast, in the backgrounds of boson stars and neutron stars, the kink ultimately transfers all of its energy to the background through oscillations, and eventually the kink disappears into the background \cite{caputo2024radial,ma2025radial}.

Of course, the present work provides a small step in this line of research. The wormhole background considered here is symmetric, however, the evolution of kinks in asymmetric wormholes deserves to be studied in future work. Our work can also be extended to axially symmetric wormholes, in order to explore the influence of angular momentum on kink dynamics. Finally, we emphasize one nontrivial observational feature revealed in this study: when the radial kink traverses the wormhole throat, it emits wave packets and radiates energy into the background in a regular manner. This phenomenon may be useful for observational searches of wormholes and for distinguishing wormholes from other compact stellar objects.

\section*{Acknowledgements}
This work was partially supported by the National Natural Science Foundation of China (Grants No.12175008).

\normalem
\bibliographystyle{ieeetr}
\bibliography{ref1.bib}

\end{document}